\documentclass[aps,pra,twocolumn,showpacs,superscriptaddress,floatfix,amsmath,amssymb]{revtex4}

\usepackage{txfonts}
\usepackage{microtype}
\usepackage{ulem}
\usepackage{amsmath}
\usepackage{amssymb}
\usepackage{bm}
\usepackage{graphicx}
\usepackage{epsfig}
\usepackage{color}

\begin{document}
\title{Momentum partition between constituents of exotic atoms during laser\\ induced tunneling ionization }

\author{Dario Cricchio}
\affiliation{Max-Planck-Institut f\"ur Kernphysik, Saupfercheckweg 1, 69117 Heidelberg, Germany}
\affiliation{Dipartimento di Fisica e Chimica, Universit\`{a} di Palermo, Via Archirafi 36, 90123 Palermo, Italy}

\author{Emilio Fiordilino}
\affiliation{Dipartimento di Fisica e Chimica, Universit\`{a} di Palermo, Via Archirafi 36, 90123 Palermo, Italy}

\author{Karen Z. Hatsagortsyan}
\affiliation{Max-Planck-Institut f\"ur Kernphysik, Saupfercheckweg 1, 69117 Heidelberg, Germany}
\thanks{\mbox{Corresponding author: k.hatsagortsyan@mpi-k.de }}

\date{\today}

\begin{abstract}

The tunneling ionization of exotic atoms such as muonic hydrogen, muonium and positronium in a strong laser field of circular polarization is investigated taking into account the impact of the motion of the center of mass on the the tunneling ionization dynamics. The momentum partition between the ionization products is deduced. The effect of the center of mass motion for
the momentum distribution of the ionization components is determined. The effect scales with the ratio of the electron (muon) to the atomic core masses and is nonnegligible for exotic atoms, while being insignificant for common atoms. It is shown that  the electron (muon) momentum shift during the under-the-barrier motion due to the magnetically induced Lorentz force has a significant impact on the momentum distribution of the atomic core and depends on the ratio of the electron to the atomic core masses.

\end{abstract}


\maketitle

\section{Introduction}

In a strong laser field  ionization of an atom takes place by absorption of multiple laser photons \cite{Protopapas_1997,Delone_Krainov_book,Becker_2002}. The photons carry momentum which is distributed among the ionized electron and the atomic core (ion) after the interaction. The photon momentum transfer in ionization is a nondipole effect,  theoretically described by a first relativistic correction to the nonrelativistic Hamiltonian. In classical terms the momentum transfer to the electron along the laser propagation direction arises due to the magnetically induced Lorentz force. At rather strong laser fields the ionization is in the tunneling regime, when the Keldysh parameter $\gamma$ is small, and the electron is released from the atom by means of tunneling through the potential barrier formed by the laser suppressed atomic potential. In the tunneling regime the Lorentz force induces the momentum transfer to the electron along the laser propagation direction either when the electron moves in continuum after releasing from the atom, or during the tunneling step of the ionization. The first effect is characterized by the the relativistic field parameter $\xi\equiv eE_0/m_ec\omega$, where the $E_0$ and $\omega$ are the laser field amplitude and angular frequency, respectively,  $-e$ and $m_e$ are the electron charge and mass, respectively, and $c$ is the speed of light. The momentum transfer along the laser propagation direction in continuum is $\Delta p_z\sim U_p/c$, where $U_p=m_ec^2\xi^2/2$ is the electron ponderomotive energy in the laser field. It is responsible for shifting of the angular distribution of photoelectrons from the laser polarization direction into the propagation direction  in relativistically strong laser fields when $\xi \sim 1$ \cite{Becker_2002,Salamin_2006}, for the suppression of the electron rescattering with the ion \cite{RMP_2012} and, consequently, suppression of nonsequential double ionization \cite{Dammasch_2001,Klaiber_2005} and high-order harmonic generation \cite{Walser_2000a,Klaiber_2007}.

The relativistic theory for the under-the-barrier dynamics demonstrated \cite{Klaiber_2013c} that the magnetically induced Lorentz force brings about a momentum shift along the laser propagation direction also during the under-the-barrier motion.  It has a characteristic  value of $I_p/c$, which can be estimated from $\Delta p_z\sim e(v_a/c)B_0\tau_k\sim I_p/c$, with the atomic velocity $v_a=\sqrt{2I_p/m_e}$, the ionization potential $I_p$, the laser magnetic field $B_0=E_0$, the Keldysh time $\tau_K=\gamma/\omega$,  and the Keldysh parameter $\gamma=\sqrt{I_p/2U_p}$.

At nonrelativistic intensities the photon momentum transfer is rather small, nevertheless, it has been measured in a recent remarkable experiment \cite{Smeenk_2011}, using nonrelativistic laser intensities of $2-10\times 10^{14}$ W/cm$^2$ and detecting electron momenta of an order of $10^{-3}$ a.u..
The experimental results \cite{Smeenk_2011} in a circularly polarized laser field indicated that after tunneling ionization the ion carries a momentum $I_p/c$ along the momentum of incoming photons, while the electron carries a momentum of $U_p/c$.  This is in accordance with calculations of \cite{Titi_2012} and with the so-called simpleman model \cite{Simpleman}. The latter assumes that the electron appears at the ionization tunnel exit with a vanishing momentum and then is driven solely by the laser field. The ion momentum in this case follows from the momentum conservation law.

However,  due to the Lorentz force effect during the under-the-barrier dynamics, the electron appears at the tunnel exit with a nonvanishing momentum along the laser propagation direction \cite{Klaiber_2013c}. The latter contributes to the asymptotic electron momentum because of which the final ion momentum decreases with respect to the prediction of the simpleman model. Thus, the peak of the electron momentum distribution along the laser propagation direction is at $p_{ez}=I_p/3c$ (if the Coulomb field of the atomic core is neglected during the under-the-barrier motion), in contrast to the simpleman model vanishing prediction, and the ion should carry a momentum $p_{iz}=2I_p/3c$ \cite{Klaiber_2013c,Chelkowski_2014}. Moreover, the ion momentum increases when the Coulomb field of the atomic core is accounted for in the near the over-the-barrier ionization regime, e.g., it is $p_{iz}=0.8I_p/c$ at $E_0/E_a=0.05$, see Fig. 6 in \cite{Yakaboylu_2013}, where $E_a=(2I_p)^{3/2}$ is the atomic field strength. As it is noted in \cite{Chelkowski_2014},  these results are not inconsistent with  the experiment   of Ref. \cite{Smeenk_2011} because of the large experimental error.
The experiment on the momentum partition at ionization is most clearly shown in a circularly polarized laser field because in this case the recollisions are avoided. In a linearly polarized field the Coulomb focusing because of recollisions  modifies significantly the final momentum distribution \cite{Comtois_2005} and complicates the analyses \cite{Ludwig_2014}. Recently, a numerical relativistic calculation of the electron momentum distribution in a linearly polarized laser field \cite{Ivanov_2015} indicated that the momentum shift in the laser propagation direction  depends linearly on the laser intensity, which is in accordance with the experiment \cite{Smeenk_2011}. An additional momentum shift can arise also because of the, so-called, tunneling time \cite{Wigner_1955,Landauer_1994,Landsman_2015} but this momentum shift is in the direction transverse to the laser propagation direction \cite{Eckle_2008a,Eckle_2008b,Landsman_2014o,Yakaboylu_2013,Yakaboylu_2014b} and will not be discussed in the present paper.

If the atomic degree of freedom is neglected during tunneling ionization of the electron, the momentum transfer to the ion can be simply deduced via the momentum conservation law, taking into account the electron momentum after the ionization. In treating the electron dynamics in strong field ionization, usually, the ion is assumed to be not moving. While it is a good approximation for the ionization of hydrogen and other common atoms because of smallness of the mass ratio of the electron to the ion, it is not the case for exotic atoms such as muonium  [the bound state of  an electron and antimuon, the mass ratio of an electron to muon is $m_e/m_{\mu}\approx 1/207$], muonic hydrogen [the bound state of a muon and a proton, the mass ratio of a  muon to proton is $ m_{\mu}/m_p \approx 0.1126$] and positronium [the bound state of the electron and positron, the mass ratio of an electron to a positron is $m_{e-}/m_{e+}= 1$]. The hydrogen ionization problem taking into account the ion degree of freedom first was considered in \cite{Chelkowski_2014} for derivation of the ion momentum, however, neglecting its impact on the electron momentum distribution.

Note that the influence of the nuclear degrees of freedom on the electronic dynamics is well known in strong field molecular processes as, so-called, non-Born-Oppenheimer dynamics, see e.g. \cite{Zhou_2012,Silva_2013,Tolstikhin_2013}.

In this paper we investigate the influence of the atomic core (ion) degree of freedom  on the electron (muon) tunneling dynamics in a strong laser field of a circular polarization in the case of hydrogen, muonium, muonic hydrogen, and positronium atoms. The impact of this effect on the photo-electron (-muon) momentum distribution and, respectively, on the partition of the photon momentum transfer between the electron (muon) and the atomic core (proton, antimuon, positron) are studied.

We label by ``electron" the negative component of the exotic atom, which is the electron in the case of muonium and positronium, and muon in the case of muonic hydrogen. The label ``ion" is employed for the positive component of the exotic atom, which is the proton in the case of muonic hydrogen, antimuon in the case of muonium, and the positron in the case of positronium.

The structure of the paper is the following. In Sec. \ref{simpleman} the momentum partition is analysed using the simpleman model along with the energy-momentum conservation law. In Sec. \ref{SFA} the result of the strong field approximation (SFA) is presented, which is followed by the discussion in Sec. \ref{discuss} and the conclusion in Sec. \ref{conclusion}.

 \section{The simpleman model and the energy-momentum conservation}\label{simpleman}

In this section we derive the  momentum partition between the electron and ion during tunneling ionization, assuming that the photoelectron dynamics follows the simpleman model \cite{Simpleman}. The information provided by energy and momentum conservation laws is used to deduce the ion momentum $\textbf{p}_i$ and the number of absorbed laser photons $n$ corresponding to a certain photoelectron momentum $\textbf{p}_e$ in the tunneling ionization process.

According to the simpleman model, the electron  appears in continuum with a vanishing momentum in the most probable trajectory and  afterwards is driven by the laser field, absorbing $n$ laser photons. The energy conservation for the ionization process reads:
\begin{equation}
  (\varepsilon_e-m_e c^2)+(\varepsilon_i-m_ic^2)=n\omega-I_p ,\label{energy}
\end{equation}
where $\varepsilon_{e,i}=c\sqrt{\textbf{p}_{e,i}^2+m_{e,i}^2c^2}$ and $m_{e,i}$ are the energy and mass of the electron and the ion, respectively. The momentum conservation provides
\begin{eqnarray}
    \textbf{p}_{e\,\bot}+\textbf{p}_{i\,\bot}&=&0 \label{ptr}\\
        p_{e\,z}+p_{i\,z}&=&\frac{n\omega}{c}, \label{Pz}
   \end{eqnarray}
where $p_{e,i\,z}$ and $ \textbf{p}_{e,i\,\bot}$ are the electron and the ion momentum components along the laser propagation direction $z$ and the transverse direction, respectively.
Combining Eqs. (\ref{energy}) and (\ref{Pz}) we derive
\begin{eqnarray}
    (\varepsilon_{e}-cp_{e\,z}-m_ec^2)+(\varepsilon_{i}-cp_{i\,z}-m_ic^2)=-I_p.
    \label{lambda}
\end{eqnarray}
In the case of a common atom ($m_i\gg m_e$) the kinetic energy of the ion is small and can be neglected, $\varepsilon_i-m_ic^2\approx 0$. In the plane laser field
\begin{equation}
\Lambda\equiv \varepsilon -cp_z
\end{equation}
is an integral of motion  \cite{RMP_2012}.  In the simpleman model $\Lambda_{e}=m_{e}c^2$ for the electron  because of vanishing of the electron kinetic momentum  at the tunnel exit. Consequently, the ion momentum is derived
\begin{eqnarray}
    p_{iz} =\frac{I_p}{c}.\label{Piz}
   \end{eqnarray}
In the simpleman model the electron transverse kinetic momentum in continuum is derived from the conservation of the transverse canonical momentum $ \textbf{p}_{e\,\bot}-e\textbf{A}(\phi)=\rm const$. Taking into account that at the ionization moment $\phi_0$ the electron kinetic momentum is vanishing, we have
\begin{eqnarray}
    \textbf{p}_{e\,\bot}=-e\left[ \textbf{A}(\phi_0)-\textbf{A}(\phi)\right],\label{p_tra}
   \end{eqnarray}
where $\phi=\omega(t-z/c)$ is the laser phase. The electron longitudinal momentum is derived  from the conservation law
\begin{equation}
\Lambda_{e,i}+cp_{e,i\,z}=c\sqrt{p_{e,i}^2+m_{e,i}^2c^2},
\end{equation}
which yields
 \begin{eqnarray}
  \label{pz_lambda}
      p_{ e,i\,z} &=&\frac{c^2p_{\bot}^2 +m_{e,i}^2c^4-\Lambda_{e,i}^2}{2c\Lambda_{e,i}}\, .
\end{eqnarray}
Assuming a vanishing transverse momentum at the tunnel exit ($\Lambda_e=m_ec^2$) yields
\begin{eqnarray}
    p_{e\,z}=\frac{e^2}{2m_e}\left[\textbf{A}(\phi_0)-\textbf{A}(\phi)\right]^2.
   \end{eqnarray}
At switching off the laser field $\textbf{A}(\phi)\rightarrow 0$:
 \begin{eqnarray}
    \textbf{p}_{e\,\bot}&=&-e \textbf{A}(\phi_0) ,\label{Pex}\\
   p_{e\,z}&=& \frac{U_p}{c},\label{Pez}\\
     \varepsilon_{e}&=&m_ec^2+U_p \label{Ee}\,,
     \end{eqnarray}
where $U_p= e^2\textbf{A}^2(\phi_0)/(2m_e)=m_ec^2\xi^2/2$. Therefore, according to the simpleman model, Eqs. (\ref{Pz}), (\ref{Piz}) and (\ref{Pez}), from the totally absorbed photon momentum during ionization
 \begin{eqnarray}
     \frac{n\omega}{c}=\frac{I_p}{c}+\frac{U_p}{c}
       \end{eqnarray}
the ion absorbs the momentum $p_{iz}=I_p/c$ and the rest of the photon momentum $U_p/c$ is transferred to the electron.

The experiment of Ref. \cite{Smeenk_2011} is in accordance with the above mentioned results of the simpleman model. However, the ion dynamics during the tunneling and the relativistic features of the electron under-the-barrier dynamics \cite{Klaiber_2013c,Chelkowski_2014,Yakaboylu_2013} have impact on the momentum partition in strong field ionization which will be discussed in the next sections. Here we shortly mention that when going beyond the simpleman model, the Lorentz force during the  under-the-barrier motion induces a nonvanishing momentum component for the electron, $p_{ez}^{(0)}$, along the laser propagation direction at the tunnel exit. Then, $\Lambda_e\approx m_ec^2-cp_{ez}^{(0)}$, when $p_{ez}^{(0)}\ll m_e c$ as usually is the case, and  from Eq.~(\ref{lambda}) one derives
\begin{eqnarray}
 p_{iz}=I_p/c-p_{ez}^{(0)}\,.\label{piz0}
\end{eqnarray}

In the case of exotic atoms, when the constituents masses are  of the same order, the simpleman condition for the vanishing momentum at the tunnel exit concerns the relative momentum
\begin{eqnarray}
  \textbf{p}^{(0)}=(m_i\textbf{p}_e^{(0)}-m_e\textbf{p}_i^{(0)})/M=0\,,
\end{eqnarray}
see Table I. Then, $\textbf{p}_{e\bot}^{(0)}=\textbf{p}_{i\bot}^{(0)}=0$, because of Eq.~(\ref{ptr}). Moreover, taking into account that  $p_{i\,z}^{(0)}=(m_i/m_e)p_{e\,z}^{(0)}$, one has
 \begin{eqnarray}
   \Lambda_i-m_ic^2=\frac{m_i}{m_e}(\Lambda_e-m_ec^2),
  \end{eqnarray}
and from Eq.~(\ref{lambda})  one derives
 \begin{eqnarray}
   \Lambda_e=m_ec^2\left(1-\frac{I_p}{Mc^2} \right),\\
     \Lambda_i=m_ic^2\left(1-\frac{I_p}{Mc^2} \right),
\end{eqnarray}
where $M=m_e+m_i$. The electron and ion momenta can be calculated from Eqs.~(\ref{pz_lambda})  and (\ref{Pex}),
which finally leads to the following expressions in the leading order of $I_p/(Mc^2)$:
\begin{eqnarray}
\textbf{p}_{ e\bot}&=&-e\textbf{A}(\phi_0)\label{p1}\,,\\
p_{e\, z}& \approx & m_ec\left[ \frac{\xi^2}{2} \left(1+\frac{I_p}{Mc^2}\right)+\frac{I_p}{Mc^2}\right]\label{p2} \, ,  \\
\textbf{p}_{ i\bot}&=& \,\,e\textbf{A}(\phi_0)\label{p1i}\,,\\
     p_{i\,z} & \approx & m_ic\left[ \frac{\xi^2}{2} \frac{m_e^2}{m_i^2}\left(1+\frac{I_p}{Mc^2}\right)+\frac{I_p}{Mc^2}\right] . \label{p3}
\end{eqnarray}
This general expression for the final electron and ion momenta according to the simpleman model includes the limiting cases of the infinitely heavy ionic core [$m_i \rightarrow \infty$, see Eqs.~(\ref{Piz}),  (\ref{Pex}) and (\ref{Pez})] and of the positronium atom [$m_e=m_i$]. In the latter case there is symmetry
between the electron and positron dynamics stemming from a positronium atom
 \begin{eqnarray}
   p_{e\,z}=p_{i\,z} \approx \dfrac{m_ec\xi^2 }{2}  \left( 1+\dfrac{I_p}{2m_e c^2}\right) +\dfrac{I_p}{2c}  \, .  \label{Ps_z}
\end{eqnarray}
where $p_{iz}$ denotes the positron momentum.

Thus, the simpleman model along with the energy-momentum conservation  provides information on momentum partition between the  electron and ion in the tunneling ionization process, given by Eqs.~(\ref{p1})-(\ref{p3}). We underline that the simpleman model does not take into account the initial momentum of the electron at the tunnel exit, which can arise due to the Lorentz force \cite{Klaiber_2013c} and due to nonadiabatic dynamics  (in the intermediate regime between the tunneling and mutiphoton regimes) \cite{Klaiber_2015}. Moreover, the Lorentz force effect depends on the ionic recoil inducing corrections of the order of $m_e/m_i$. Note also that the final momentum distribution is disturbed also by the Coulomb focusing during the electron motion in the continuum \cite{Comtois_2005}.

In the next sections the Lorentz force effect and the impact of the ion recoil for the momentum partitioning between the ion and the electron are discussed.

\section{Strong field approximation}\label{SFA}

We consider strong field ionization of a simple atomic system consisting of a positively and a negatively charged particles (labelled as "ion" and "electron", respectively) which are initially in the ground state of the bound system. The main aim is to study the influence of the ion motion on the electron tunneling dynamics and its impact on the momentum partitioning between the ionized electron and the atomic core in tunnelling ionization. The theory  will be applied for the cases of ionization of hydrogen, muonium, muonic hydrogen, and for positronium. The degree of freedom both of the ion and the electron will be taken into account. The effect of the magnetically induced Lorentz force on the ionization dynamics is included by means of nondipole treatment of the laser field in the weakly relativistic regime.

\begin{table}[b]
\begin{tabular}{|c|c|c|}
\hline
$\mathbf{R}=(X,Y,Z)=\dfrac{m_i\mathbf{r}_i+m_e\mathbf{r}_e}{m_i+m_e}$ & $\textbf{r}_i = \textbf{R}-\dfrac{m_e}{M}\textbf{r}$ & $t-\dfrac{z_i}{c}=\tau+\dfrac{m_e}{Mc}z$ \\
\hline
$\mathbf{r}=(x,y,z)=\mathbf{r}_e-\mathbf{r}_i$ & $\textbf{r}_e=  \textbf{R}+\dfrac{m_i}{M}\textbf{r}$ & $t-\dfrac{z_e}{c}=\tau-\dfrac{m_i}{Mc}z$  \\
\hline
$\mu=\dfrac{m_im_e}{m_i+m_e}$ & $\boldsymbol{\nabla}_i=\dfrac{m_i}{M}\boldsymbol{\nabla}_R-\boldsymbol{\nabla}_r$ & $\tau=t-\dfrac{Z}{c}$ \\
\hline
$M=m_i+m_e$ & $\boldsymbol{\nabla}_e=\dfrac{m_e}{M}\boldsymbol{\nabla}_R+\boldsymbol{\nabla}_r$ & $\zeta=Z$\\
\hline
$\eta\equiv\dfrac{m_i^2-m_e^2}{m_im_e}$ & $\textbf{P}=\textbf{p}_e+\textbf{p}_i $  & $\dfrac{\partial}{\partial Z}=\dfrac{\partial}{\partial \zeta}- \dfrac{\partial}{c\partial \tau}$\\
\hline
$\hat{\textbf{P}}=-i\boldsymbol{\nabla}_R$, $\hat{\textbf{p}}=-i\boldsymbol{\nabla}$, $\dot{\textbf{p}}=\mu\dot{\textbf{r}}$ &$\textbf{p}=\dfrac{m_i\textbf{p}_e-m_e\textbf{p}_i}{M} $&$\dfrac{\partial}{\partial t}=\dfrac{\partial}{\partial \tau}$\\
\hline
\end{tabular}
\label{tab:conv}
\caption{The variable transformation to the relative and  c.m. coordinates and, further, to the light-time.}
\end{table}

The Hamiltonian of the system under the consideration, consisting of an ion and an electron in a laser field reads:
\begin{eqnarray}
\mathcal{H}&=&\dfrac{1}{2m_i}\Bigg[-i\boldsymbol{\nabla}_i-q_i\mathbf{A}\left(t-\dfrac{z_i}{c}\right)\Bigg]^2\\
&+&\dfrac{1}{2m_e}\Bigg[-i\boldsymbol{\nabla}_e-q_e\mathbf{A}\left(t-\dfrac{z_e}{c}\right)\Bigg]^2+\dfrac{q_iq_e}{\mid \mathbf{r}_e-\mathbf{r}_i\mid}\nonumber,
\end{eqnarray}
where $\textbf{r}_{e,i}$, $q_i=1$ and $q_e=-1$  are the radius-vectors and charges of the ion and the electron, respectively (henceforth atomic units are used). To simplify the calculations it is convenient to write the Hamiltonian in the relative and the center-of-mass (c.m.) coordinates:
\begin{eqnarray}
\mathcal{H}&=&\dfrac{1}{2m_i}\Bigg[-i\dfrac{m_i}{M}\boldsymbol{\nabla}_R+i\boldsymbol{\nabla}_r-q_i\mathbf{A}\left(\tau+\dfrac{m_e}{M}\dfrac{z}{c}\right)\Bigg]^2\\
&+&\dfrac{1}{2m_e}\Bigg[-i\dfrac{m_e}{M}\boldsymbol{\nabla}_R-i\boldsymbol{\nabla}_r-q_e\mathbf{A}\left(\tau-\dfrac{m_i}{M}\dfrac{z}{c}\right)\Bigg]^2+\dfrac{q_iq_e}{r}\nonumber
\end{eqnarray}
The variable transformation is shown in Table I.
We use nondipole  description in order to study the dynamic of the system under the influence of the magnetic component of the laser field which is responsible for the momentum transfer from the photons to the electron and the ion. The circularly polarized laser field propagating in $z$-direction is described by a vector potential:
\begin{eqnarray}
A_x (z_\alpha,t)&=& A_0\cos \omega( t-z_\alpha/c) \nonumber\\
A_y(z_\alpha,t) &=& A_0\sin \omega( t-z_\alpha /c)
\end{eqnarray}
with $A_0=E_0/\omega$ and $\alpha\in\{i,e\}$.

The ionization transition amplitude is calculated using SFA \cite{Keldysh_1965,Faisal_1973,Reiss_1980}:
\begin{eqnarray}
M_{fi}&=&-i\int_{-\infty}^{\infty}\textbf{d}t\,\langle \Psi_{\textbf{P},\textbf{p}}(t)\mid V_L(t)\mid \Psi_{0,\textbf{P}_0}(t)\rangle \label{Mfi}
\end{eqnarray}
where
\begin{eqnarray}
V_L(t)\equiv \sum_\alpha \left[-\frac{q_\alpha}{m_e}\textbf{p}_\alpha\textbf{A}(z_\alpha,t)+\frac{q_\alpha^2\textbf{A}^2(z_\alpha,t)}{2m_\alpha}\right] \nonumber
\end{eqnarray}
describes the interaction with the laser field; $|\Psi_{0,\textbf{P}_{0}}(t)\rangle=|\Phi_0 \rangle e^{iI_pt+i\textbf{P} _{0}\textbf{R}}$ is the initial bound state of the electron-ion system which is in the ground state $|\Phi_0 \rangle$ with the energy $-I_p$; the momentum of c.m. of the electron-ion system is $\textbf{P}_{0}$; $|\Psi_{\textbf{P} ,\textbf{p} } (t)\rangle $ is the continuum state of the electron and ion in the laser field  with the asymptotic c.m. momentum $\textbf{P} $ and the relative momentum $\textbf{p} $, neglecting Coulomb interaction.

Similar to \cite{Becker_2002}, the transition matrix element of Eq.~(\ref{Mfi}) can be represented as
 \begin{eqnarray}
M_{fi}&=&-i\int_{-\infty}^{\infty}\textbf{d}t\,\langle  \Psi_{\textbf{P} ,\textbf{p} }(t)\mid V(\textbf{r})\mid \Psi_{0,\textbf{P} _{0}}(t) \rangle,
\end{eqnarray}
where $V(\textbf{r})=q_iq_e/r$ is the atomic potential.

The continuum wave function for the electron and ion system in the laser field, after transformation to the light-time  $\tau =t-Z/c$, where $Z$ is the c.m. coordinate along the laser propagation direction, fulfils the equation
\begin{equation}
 i \partial_\tau  \Psi_{\textbf{P} ,\textbf{p} }(\tau) = \hat{H }\Psi_{\textbf{P} ,\textbf{p} }(\tau)\, , \label{Volkov}
 \end{equation}
with
 \begin{eqnarray}
 \hat{H }= \dfrac{1}{2m_i}\Bigg[-i\dfrac{m_i}{M}\left(\boldsymbol{\nabla}_R -\frac{\hat{\textbf{z}}}{c}\partial_\tau\right)+i\boldsymbol{\nabla}-q_i\mathbf{A}\left(\tau+\dfrac{m_e}{M}\dfrac{z}{c}\right)\Bigg]^2 \nonumber\\
 + \dfrac{1}{2m_e}\Bigg[-i\dfrac{m_e}{M}\left(\boldsymbol{\nabla}_R-\frac{\hat{\textbf{z}}}{c}\partial_\tau\right)-i\boldsymbol{\nabla}-q_e\mathbf{A}\left(\tau-\dfrac{m_i}{M}\dfrac{z}{c}\right)\Bigg]^2 \,.\nonumber
\end{eqnarray}
Taking into account the conservation law $[\hat{\textbf{P}},\hat{H}]=0$, the c.m. coordinates are factorized in the wave function:
\begin{equation}
  \Psi_{\textbf{P} ,\textbf{p} }(\tau,\textbf{r},\textbf{R})=\exp\left(i\textbf{P}\cdot \textbf{R} -i{\cal E} t\right)\phi(\textbf{r},\tau),
 \end{equation}
with the c.m. energy ${\cal E}$ and momentum  $\textbf{P}$, after the interaction is switched off.

In the weakly-relativistic regime  the vector potential can be expanded to the first order in $1/c$:
\begin{eqnarray}
\mathbf{\textbf{A}}\left(\tau\pm \dfrac{m_\alpha z}{Mc}\right)\approx \mathbf{A}(\tau)\mp\dfrac{m_\alpha z}{Mc} \mathbf{E}(\tau).
\end{eqnarray}
Then, neglecting the high-order terms over $1/c$ in Eq.~(\ref{Volkov}), we arrive at the following equation:
\begin{eqnarray}
 &&\left\{ i\left( 1-\dfrac{P_{z}}{Mc} \right)\partial_\tau -\frac{1}{2\mu}\left[-i\boldsymbol{\nabla}+ \textbf{A}(\tau)\right]^2 \right.\\
 &+&\left.\dfrac{z}{m_ec}\left[-i\boldsymbol{\nabla}+ \textbf{A}(\tau)\right]\textbf{A}'(\tau) \left(1-\dfrac{m_e}{m_i} \right)\right.\nonumber\\
 &-&\left.\dfrac{P^2}{2M}+\dfrac{z}{Mc}\textbf{P}\cdot\textbf{A}'(\tau)+{\cal E}\right\}\phi (\textbf{r},\tau)=0\,.\nonumber
\end{eqnarray}
When the wave function is parametrized as
\begin{eqnarray}
\phi (\textbf{r},\tau)=\exp[i\textbf{p}\cdot \textbf{r}+iT(\tau)z-iS(\tau)]\,,
\end{eqnarray}
one obtains
\begin{eqnarray}
 T(\tau)&=&\frac{\eta}{Mc\left(1-\frac{P_z}{Mc}\right)} \int_{-\infty}^\tau d\tau' \left[ \textbf{p}+\textbf{A}(\tau')\right]  \textbf{A}'(\tau')\,,\\
 S(\tau)&=&\frac{1}{ 2\mu\left(1-\frac{P_z}{Mc}\right)}\int_{-\infty}^\tau d\tau' \left[ \textbf{p}+\hat{\textbf{z}}T(\tau')+\textbf{A}(\tau')\right]^2 \,,
\end{eqnarray}
where $\mu$ and $\eta$ are defined in Table I. Finally, the continuum wave function for the electron-ion system in the laser field in the leading order of the $1/c$-expansion is
\begin{equation}
\Psi_{\textbf{P} ,\textbf{p} }(\tau,\textbf{r},\textbf{R})={\cal N}\exp \left(i{\cal S} \right) \, ,
\end{equation}
with the normalization constant ${\cal N}$ and the action
\begin{eqnarray}
{\cal S}   &=&\mathbf{P} \cdot\mathbf{R}-{\cal E} t+\mathbf{p}\cdot\mathbf{r}+\dfrac{\eta \,z}{Mc}\left[\mathbf{p}\cdot\mathbf{A}(\tau)+\dfrac{\mathbf{A}^2(\tau)}{2}\right] \nonumber\\
&-&\frac{1}{\mu}\left(1+\dfrac{P_{z}+\eta p_{z}}{Mc}\right)\int_{-\infty}^\tau d\tau'\left[\mathbf{p}\cdot\mathbf{A}(\tau')+\dfrac{\mathbf{A}^2(\tau')}{2}\right]\,. \label{SSS}
\end{eqnarray}
We calculate the matrix elements of the amplitude between the ground state and the continuum as:
\begin{eqnarray}
M_{fi}&=&-i{\cal N}\int\textbf{d} t \int\textbf{d}\mathbf{R}\int\textbf{d}\mathbf{r}\,e^{-i{\cal S}}V(r)\Phi_0(r)e^{iI_pt},
\end{eqnarray}
assuming that the atom is at rest in the initial state $\mathbf{P}_0=0$.

Let us first consider the momentum sharing between the electron and the ion in the simplest and transparent case  when the atomic potential is modelled  by a short-range potential. Later we will discuss the correction to this picture due to the real atomic potential. In the case of short-range potential \cite{Bersons_1975}
\begin{equation}
V(r)=(2\pi/\kappa)\delta (\textbf{r})\partial_r r,\label{short-range}
\end{equation}
one has $\langle \textbf{p}|V|\Phi_0\rangle =-\sqrt{\kappa}/(2\pi)$, with $\kappa\equiv \sqrt{2\mu I_p}$.
We expand the last term of the action in Eq.~(\ref{SSS}):
\begin{eqnarray}
&&\exp\left\{-\dfrac{i}{\mu}\left(1+\dfrac{P_{z}+\eta p_{z}}{Mc}\right)\int_{-\infty}^\tau \left(\mathbf{p}\cdot\mathbf{A}(\tau')+\dfrac{A^2}{2}\right)\textbf{d}\tau'\right\}\nonumber\\
 &=&\sum_{n=-\infty}^{\infty}i^nJ_n(\zeta)\exp\left\{in\omega\tau+i\sigma_0\tau+in\varphi_0\right\}\,,
\end{eqnarray}
where $J_n(\zeta)$ is the Bessel functions, $n$ is the number of absorbed photons, $\tan\varphi_0=p_x/p_y$, and
\begin{eqnarray}
\zeta &=& \left(1+\frac{P_{z}+\eta p_{z}}{Mc}\right)\frac{p_{\bot}A_0}{\mu\omega},\\
 \sigma_0 &=& \left(1+\frac{P_{z}+\eta p_{z}}{Mc}\right)\frac{A_0^2}{2\mu},
\end{eqnarray}
and obtain for the transition amplitude
\begin{eqnarray}
M_{fi}&=&i(2\pi)^3{\cal N}\sqrt{\kappa }\,\delta\left(\mathbf{P}_{\perp}\right)\sum_{n=-\infty}^{\infty}\delta\left(P_{Z}+\dfrac{\sigma_0}{c}-\dfrac{n\omega}{c}\right)\nonumber\\
&\times &
\delta\left(\dfrac{P_{Z}^2}{2M}+\dfrac{p^2}{2\mu}+I_p+\sigma_0-n\omega\right)J_n\left(\zeta\right)e^{i n\varphi_0}
\end{eqnarray}
Then we can calculate the ionization rate
\begin{equation}
dW=  |M_{fi}|^2 \frac{d^3\textbf{P}}{(2\pi)^3}\frac{d^3\textbf{p}}{(2\pi)^3}= |M_{fi}|^2  \frac{d^3\textbf{p}_i}{(2\pi)^3}\frac{d^3\textbf{p}_e}{(2\pi)^3}\, .
\end{equation}
Describing the final phase space via the electron and ion momenta, the ionization rate reads
\begin{eqnarray}
\frac{dW}{d^3\textbf{p}_id^3\textbf{p}_e}&=&\frac{\kappa}{(2\pi)^4} \sum_n \delta \left( \textbf{p}_{e\bot}+\textbf{p}_{i\bot}\right) \delta \left( p_{ez}+p_{iz}-\frac{n\omega-\sigma_0}{c}\right) \nonumber\\
&\times& \delta \left(\Delta\right) \,J_n^2(\zeta)\,, \label{DRate}
\end{eqnarray}
where
\begin{eqnarray}
\Delta &\equiv & I_p-n\omega+\sigma_0+\dfrac{p_e^2}{2m_e} +\dfrac{p_i^2}{2m_i}\, , \nonumber\\
\sigma_0 &=& \dfrac{m_e^2c^2\xi^2}{2\mu}\left[1+ \dfrac{p_{ez}+p_{iz}}{Mc}+\frac{\eta}{Mc}\left(\dfrac{m_i}{M}p_{ez}- \dfrac{m_e}{M}p_{iz}\right)\right]\, ,\nonumber\\
\zeta &=&\dfrac{m_ec \xi p_{e\perp}}{\mu\omega}\left[1+ \dfrac{p_{ez}+p_{iz}}{Mc}+\frac{\eta}{Mc}\left(\dfrac{m_i}{M}p_{ez}- \frac{m_e}{M}p_{iz}\right)\right]\, .\nonumber
\end{eqnarray}
After the integration over the ion momenta, one obtains for the photoelectron momentum distribution \begin{equation}
\frac{dW}{d^3\mathbf{p}_e}=\frac{\kappa}{(2\pi)^4} \dfrac{1}{1+\frac{\xi^2}{2}\frac{m_e^2}{m_i^2}}\sum_n J_n^2(\zeta)\delta\left(\Delta\right)\, .\label{Rate}
\end{equation}
Here we take into account that
\begin{equation}
\left|\dfrac{\partial}{\partial p_{iz}} p_{ez}+p_{iz} \dfrac{n\omega-\sigma_0}{c}\right|=1+\dfrac{\xi^2}{2}\dfrac{m_e^2}{m_i^2}. \nonumber
\end{equation}
Expressing the ion momentum via the electron momentum using the momentum conservation $\delta$-function, we have \begin{eqnarray}
\zeta &=&\dfrac{m_ec \xi p_{e\perp}}{\mu\omega}\left[1+ \left(1-\dfrac{m_e}{m_i} \right)\dfrac{p_{ez}}{m_ec}+ \dfrac{m_e^2}{M m_i}\nu\right]\,,\label{zeta}\\
\nu &=&\frac{1}{1+\frac{\xi^2}{2}\frac{m_e^2}{m_i^2}}\left\{\dfrac{n\omega}{m_ec^2}-\dfrac{\xi^2}{2}\dfrac{m_e}{\mu}\left[1+ \left(1-\dfrac{m_e}{m_i} \right)\dfrac{p_{ez}}{m_ec}\right]\right\} \label{nu}  \,,\\
\Delta  &=&m_ec^2\left\{\tilde{I}_p+\dfrac{p_e^2}{2\mu m_ec^2}-\nu\left(1+\dfrac{p_{ez}}{m_ic} \right)+\dfrac{\nu^2}{2}\dfrac{m_e}{m_i}\right\} \, ,\label{delta}
\end{eqnarray}
where $\tilde{I}_p\equiv I_p/m_ec^2$ and $\nu \equiv (n\omega-\sigma_0)/m_ec^2$ .

The ionization rate given by Eq. (\ref{Rate}) depends on ion mass and takes into account the impact of the motion of the c.m. of the electron-ion system on the tunneling dynamics. For the hydrogen atom it is negligible as it scales with the  small ratio $m_e/m_i\approx 1/1836$. However, for exotic atoms this effect cannot be neglected.

\section{Discussion}\label{discuss}

Let us analyse the ionization differential rate  to find out the most probable momentum of the ionized electron and ion in the case of different  atomic systems. We can approximately replace the summation over the photon number $n$ in Eq.~(\ref{Rate}) by integration and carry out the latter using the $\delta$-function:
\begin{equation}
\frac{dW}{d^3\mathbf{p}_e}\approx\frac{\kappa}{(2\pi)^4}\dfrac{J_n^2(\zeta)}{\omega\left( 1+\dfrac{p_{ez}}{m_ic}-\dfrac{m_e}{m_i} \nu\right)}  \, , \label{Rate2}
\end{equation}
where we have used that
\begin{equation}
 \dfrac{\partial \Delta}{\partial n}=\frac{\omega}{1+\frac{\xi^2}{2}\frac{m_e^2}{m_i^2}}
 \left( 1+\dfrac{p_{ez}}{m_ic}-\dfrac{m_e}{m_i} \nu\right)\,.
 \end{equation}
Here, the number of absorbed photons, or the parameter $\nu$, is determined from the energy conservation $\Delta=0$, whose approximate solution  reads
\begin{equation}
 \nu\approx \dfrac{\tilde{I}_p+\dfrac{p_e^2}{2\mu m_ec^2}}{  1+\dfrac{p_{ez}}{m_ic}} \,,\label{nu2}
\end{equation}
where we have used that $\tilde{I}_p\ll 1$ and $p_e^2\ll \mu m_ec^2$.
Accordingly, the number of absorbed laser photons is
\begin{eqnarray}
 n &=& \dfrac{m_ec^2}{\omega}\left\{ \nu \left(1+\frac{\xi^2}{2}\frac{m_e^2}{m_i^2} \right)+\dfrac{\xi^2M}{2m_i} \left[1+\left(1-\dfrac{m_e}{m_i} \right) \dfrac{p_{ez}}{m_ec}\right]\right\}\,,\nonumber\\
\label{nnn}
\end{eqnarray}
The qualitative behaviour of the momentum distribution according to the differential ionization rate of Eq.~(\ref{Rate2}) is determined by the Bessel function. In the tunneling regime, when $U_p/\omega \gg 1$ and $I_p\gg 1$, one has $\zeta \gg 1$, therefore, for the further analysis we will use the asymptotics of the Bessel function $n\sim \zeta \rightarrow \infty$ \citep{Abramowitz}:
\begin{equation}
  J_n^2(\zeta)\sim \dfrac{1}{2\pi\sqrt{2(n-\zeta)\zeta}}\exp\left\{-\dfrac{4\sqrt{2}}{3}\dfrac{ (n-\zeta)^{3/2}}{\zeta^{1/2}}\right\} \,.
\end{equation}
The peak of the momentum distribution corresponds to the minimum of the expression ${\cal F}\equiv (n-\zeta)^{3}/\zeta$, achievable at $n\sim \zeta$, where $n$ and $\zeta$  are given by Eqs.~(\ref{nnn}) and (\ref{zeta}).

\subsection{Hydrogen atom}\label{Hydrogen}

First, let us consider the simplest limit of infinitely heavy ion $m_i\rightarrow \infty$.
In this case,
\begin{eqnarray}
\nu &=& \dfrac{\varkappa^2}{2}+\dfrac{p_\bot^2}{2}+\dfrac{p_z^2}{2}, \label{NU}\\
\zeta &=&\dfrac{m_ec^2}{\omega}\xi p_\bot(1+p_z), \\
n &=& \dfrac{m_ec^2}{\omega}\left[\nu+\dfrac{\xi^2(1+p_z)}{2}\right],
\end{eqnarray}
with $\varkappa^2/2\equiv\tilde{I}_p$, $p_\bot\equiv p_{e\bot}/m_ec$ and $p_z\equiv p_{e\,z}/m_ec$, which yields
\begin{equation}
 {\cal F}(p_\bot,p_z)=\left(\dfrac{m_ec^2}{\omega}\right)^2\dfrac{\left[ \varkappa^2+p_\bot^2+p_z^2+(\xi^2-2p_\bot\xi)(1+p_z)\right]^3}{8\xi p_\bot(1+p_z)}  \,.
\end{equation}
The conditions for ${\cal F}=\rm min$, $\partial{\cal F}/\partial p_\bot=0 $ and $\partial{\cal F}/\partial p_z=0 $, read, respectively:
\begin{eqnarray}
  && 6\left[p_\bot-\xi(1+p_z) \right]p_\bot = \varkappa^2+p_\bot^2+p_z^2+(\xi^2-2p_\bot\xi)(1+p_z),\nonumber \\
    && 3(1+p_z)\left[2p_z+\xi^2-2p_\bot\xi  \right]=  \\
   &&\,\,\,\,\,\,\,\,\,\,\,\,\,\,\,\,\,\,\,\,\,\,\,\,\,\,\,\,\,\,\,\,\,\,\,\,\, = \varkappa^2+p_\bot^2+p_z^2+(\xi^2-2p_\bot\xi)(1+p_z).   \nonumber
 \end{eqnarray}
Solving the latter in perturbation with respect to $\xi$ and taking into account that $p_\bot \sim \xi$ and $p_z\sim \xi^2$, as well as $\varkappa\sim \xi$, we obtain:
\begin{eqnarray}
    p_{e\bot} &=& m_ec\xi\left(1+\dfrac{\gamma^2}{6}\right) \, ,\label{Pbote}\\
    p_{e\,z} &=& \dfrac{I_p}{3c}+\dfrac{p_{e\bot}^2}{2m_ec}\, .\label{PZe}
\end{eqnarray}
In the latter  the leading order terms with respect to $\gamma$ are retained ($\gamma=\varkappa/\xi< 1$ in the tunnelling regime). The ion momentum can be deduced from the $\delta$-functions of Eq.~(\ref{DRate}),
\begin{eqnarray}
   p_{i\, z}&= &\nu m_ec^2-p_{e\, z}\, ,\label{PiZ}\\
  p_{i\, \bot}&= & -p_{e\, \bot}\, \label{PiTR}.
 \end{eqnarray}
and using Eqs. (\ref{nu}),(\ref{Pbote}), and (\ref{PZe}):
\begin{eqnarray}
     p_{i\,z} &=& \dfrac{2I_p}{3c} \, .\label{PZI}
\end{eqnarray}
Comparing Eq.~(\ref{PZI}) with our qualitative discussion in Sec. II, see Eq.~(\ref{piz0}), we can conclude that the electron momentum at the tunnel exit is $p_z^{(0)}=I_p/3c$, which is the reason of variation the ion momentum from the $I_p/c$ value. This result coincides with the predictions of Refs. \cite{Klaiber_2013c,Chelkowski_2014}.

\subsection{Exotic atoms}

In the case of exotic atoms, such as positronium ($m_e/m_i=1$), muonic hydrogen atom ($m_e/m_i\approx 0.1126$), and  muonium ($m_e/m_i\approx 1/207$), the masses of constituents are comparable and, therefore we have to use the general expressions for the parameters $n$ and $\zeta$, given by Eqs.~(\ref{zeta}),(\ref{nu}), and (\ref{nnn}).

First, we find an approximate solution of Eq. (\ref{nu}) in perturbation with respect to the parameter $\xi$:
\begin{equation}
\nu\approx \nu^{(2)}+\nu^{(4)}\,,
\end{equation}
where $\nu^{(n)}\sim \xi^n$, assuming that $p_\bot \sim \xi$ and $p_z\sim \xi^2$.
\begin{eqnarray}
\nu^{(2)} &= & \dfrac{p_\bot^2}{2}\left(1+\tilde{\mu} \right)+ \dfrac{\varkappa^2}{2}, \label{NU2}\\
 \nu^{(4)} &= & \dfrac{p_z^2}{2}\left(1+\tilde{\mu} \right)-\nu^{(2)}\tilde{\mu}p_z+\dfrac{\nu^{(2)^2}}{2}\tilde{\mu}\,,
\end{eqnarray}
where  $\tilde{\mu}\equiv m_e/m_i$. The parameters of the Bessel functions up to the order of $\xi^4$ are:
\begin{eqnarray}
\zeta &\approx&\dfrac{m_ec^2}{\omega}(1+\tilde{\mu})\xi p_\bot\left[1+(1-\tilde{\mu})p_z+\dfrac{\tilde{\mu}^2}{1+\tilde{\mu}}\nu^{(2)} \right] , \\
n &\approx & \dfrac{m_ec^2}{\omega}\left\{\nu^{(2)}\left(1+ \dfrac{\xi^2\tilde{\mu}^2}{2} \right)\right.\nonumber\\
&+& \left. \nu^{(4)}+\dfrac{\xi^2}{2}\left[1+\tilde{\mu}+(1-\tilde{\mu}^2)p_z\right]\right\}.
\end{eqnarray}
The condition $\partial {\cal F}/\partial p_\bot=0$  in this case yields:
\begin{eqnarray}
 && 3p_\bot(1+\tilde{\mu})\left[ 1+(1-\tilde{\mu})p_z+\frac{\tilde{\mu}^2}{1+\tilde{\mu}}\nu^{(2)}\right]\left\{(1+\xi^2\tilde{\mu}^2)p_\bot \right.\, \nonumber\\
 && +\left. \tilde{\mu} p_\bot(\nu^{(2)}-p_z)-\xi\left[1+ (1-\tilde{\mu})p_z+\frac{\tilde{\mu}^2}{1+\tilde{\mu}}\nu^{(2)}+p_\bot^2\tilde{\mu}^2\right]\right\} \nonumber\\
 && =\left\{\nu^{(2)}\left(1+\frac{\xi^2\tilde{\mu}^2}{2}\right) +\nu^{(4)}+(1+\tilde{\mu})\left[1+(1-\tilde{\mu})p_z \right]\frac{\xi^2}{2}\right.\nonumber\\
&& - \left.(1+\tilde{\mu})\xi p_\bot \left[1+\tilde{\mu}+(1-\tilde{\mu})p_z+\frac{\tilde{\mu}^2}{1+\tilde{\mu}}\nu^{(2)}\right]\right\}\nonumber \\
&& \times \left[1+ (1-\tilde{\mu})p_z+\frac{\tilde{\mu}^2}{1+\tilde{\mu}}\nu^{(2)}+p_\bot^2\tilde{\mu}^2\right],
\label{condition_bot}
\end{eqnarray}
which in the leading order reads
\begin{eqnarray}
 3p_\bot(p_\bot-\xi)=\frac{(p_\bot-\xi)^2}{2}+\frac{\varkappa^2}{2(1+\tilde{\mu})}.\label{condition_bot2}
\end{eqnarray}
The solution of the latter provides us the most probable transverse momentum:
\begin{eqnarray}
 p_{e\bot}=m_ec\xi\left(1+\frac{\gamma^2}{6}\dfrac{1}{1+\frac{m_e}{m_i}} \right) \, .\label{PTR}
\end{eqnarray}
The derived transverse momentum component  contains a nonadiabatic correction, the term $\sim\gamma^2$ in Eq.~(\ref{PTR}), which is absent in our simpleman estimation  via Eqs.~(\ref{p1}) and which is disturbed by the ion recoil (see the term $m_e/m_i$).

The second condition of the maximal probability $\partial {\cal F}/\partial p_z=0$ is
\begin{eqnarray}
 && 3 \left[ 1+(1-\tilde{\mu})p_z+\frac{\tilde{\mu}^2}{1+\tilde{\mu}}\nu^{(2)}\right]\left\{(1-\tilde{\mu}^2)\frac{\xi^2}{2}+p_z(1+\tilde{\mu})-\nu^{(2)}\tilde{\mu}  \right.\, \nonumber\\
 &&\left. -(1-\tilde{\mu}^2)\xi p_\bot \right\}=(1-\tilde{\mu})\left\{\nu^{(2)}\left(1+\frac{\xi^2\tilde{\mu}^2}{2}\right) +\nu^{(4)}\right. \nonumber \\
 &&+(1+\tilde{\mu})\left[1+(1-\tilde{\mu})p_z \right]\frac{\xi^2}{2} \nonumber\\
&& - \left.(1+\tilde{\mu})\xi p_\bot \left[1+\tilde{\mu}+(1-\tilde{\mu})p_z+\frac{\tilde{\mu}^2}{1+\tilde{\mu}}\nu^{(2)}\right]\right\}.
\label{condition_z}
\end{eqnarray}
In the leading order, the Eq.~(\ref{condition_z}) is simplified:
\begin{eqnarray}
&& 3 \left[(1-\tilde{\mu})\frac{\xi^2}{2}+p_z-\frac{p_\bot^2}{2}\tilde{\mu}-\tilde{\mu}\frac{\varkappa^2}{2(1+\tilde{\mu})}-(1-\tilde{\mu})\xi p_\bot \right]\nonumber\\
&& =(1-\tilde{\mu}) \left[\frac{p_\bot^2}{2}+\frac{\varkappa^2}{2(1+\tilde{\mu})}+ \frac{\xi^2}{2}-\xi p_\bot \right],
\label{condition_z}
\end{eqnarray}
the solution of which provides us the most probable longitudinal momentum:
\begin{eqnarray}
 p_{e\, z}=\frac{p_{e\bot}^2}{2m_ec}+\frac{I_p}{3c}\left(1+\frac{m_e}{M} \right)\, .\label{PeZ}
\end{eqnarray}
The ion momentum is derived from  Eqs.~(\ref{PiZ}) and (\ref{PiTR}). The second term corresponds to the momentum of the electron at the tunnel exit. The longitudinal component of the ion momentum, then, is
\begin{eqnarray}
 p_{i\, z} =\frac{p_\bot^2}{2m_ec}\frac{m_e}{m_i}+\frac{2I_p}{3c}\left( 1-\frac{m_e}{2M}\right).\label{PIZ}
\end{eqnarray}
The electron and ion longitudinal momentum Eqs.~(\ref{PeZ}) and (\ref{PIZ}) are different from the prediction of the simpleman model Eqs.~(\ref{p2}) and (\ref{p3}). It is due to the nonvanishing electron-ion relative momentum at the tunneling exit which depends on the ionization energy $I_p$ as well as on the mass ratio $m_e/M$. The latter factor describes the role of the ion motion during the tunneling process.

Now, from Eqs.~(\ref{PTR}), (\ref{PeZ}),  and (\ref{PIZ}), we are able to evaluate the most probable momentum for ionization of exotic atoms, taking into account the effect of the ion motion on the ionization dynamics.

\subsubsection{Muonium}

In the case of muonium (electron and antimuon) the most probable momentum of the electron is
 \begin{eqnarray}
 p_{e\bot}&\approx& m_ec\xi\left(1+ 0.166\gamma^2 \right) \, ,\label{mueTR}\\
 p_{e\, z}&\approx&\frac{p_{e\bot}^2}{2m_ec}+0.335\frac{I_p}{c}  \, .\label{mueZ}
\end{eqnarray}
The momentum of the antimuon is
\begin{eqnarray}
 p_{\bar{\mu}\bot}&\approx& m_ec\xi\left(1+ 0.166\gamma^2 \right) \, ,\label{mumTR}\\
 p_{\bar{\mu}\, z}&\approx&\frac{p_{e\bot}^2}{414m_{e}c}+0.665\frac{I_p}{c}  \, .\label{mumZ}
\end{eqnarray}

\subsubsection{Muonic hydrogen}

In the case of a muonic hydrogen  atom (muon and proton) the most probable momentum of the muon is
\begin{eqnarray}
 p_{\mu\bot}&\approx & m_ec\xi\left(1+0.1498\gamma^2  \right) \, ,\label{mhTR}\\
 p_{\mu\, z}&\approx&\frac{p_{\mu\bot}^2}{2m_\mu c}+0.367\frac{I_p}{c}  \, ,\label{mhZ}
\end{eqnarray}
While for the proton they are
\begin{eqnarray}
 p_{p\bot}&\approx & m_ec\xi\left(1+0.1498\gamma^2  \right) \, ,\label{mhpTR}\\
 p_{p\, z}&\approx&\frac{p_{\mu\bot}^2}{16m_\mu c}+0.633\frac{I_p}{c}  \, ,\label{mhpZ}
\end{eqnarray}

\subsubsection{Positronium}

In the case of a positronium atom the most probable momentum of the photoelectron is
\begin{eqnarray}
 p_{e\bot}&=& m_ec\xi\left(1+\frac{\gamma^2}{12} \right) \, ,\label{PsTR}\\
 p_{e\, z}&=&\frac{p_{e\bot}^2}{2m_ec}+\frac{I_p}{2c}  \, .\label{Ps_Z}
\end{eqnarray}
The positron momentum components are the same by the absolute value (the transverse momentum is opposite).

\subsection{The role of the Coulombic atomic potential}\label{Coulomb}

In the discussion above, we assumed  a short-range atomic potential, Eq.~(\ref{short-range}). Now we examine how the SFA calculations are modified when the exact Coulombic atomic potential is employed. In this case the matrix element $\langle \textbf{p}|V|\Phi_0\rangle =-\sqrt{\kappa}/(2\pi)$ should be replaced by
\begin{eqnarray}
\langle \textbf{p}|V|\Phi_0\rangle &=& \frac{4\sqrt{\pi}\alpha\kappa^{3/2}}{\kappa^2+p_\bot^2+\left[p_z+\beta(\tau)\right]^2}\nonumber\\
&\approx & \frac{4\sqrt{\pi}\alpha\kappa^{3/2}}{\kappa^2+p^2} \left[1-\dfrac{2p_z\beta(\tau)}{\kappa^2+p^2}\right],\label{V_C}
\end{eqnarray}
where $\alpha=e^2/\hbar$ and
\begin{eqnarray}
\beta(\tau)&=&\beta_0+\beta_1\sin(\omega\tau+\varphi_0)\nonumber \\
\beta_0 &\equiv & \left(1-\dfrac{m_e}{m_i} \right)\dfrac{m_ec\xi^2}{2},\nonumber \\
\beta_1 &\equiv & \left(1-\dfrac{m_e}{m_i} \right)\xi p_\bot \nonumber
\end{eqnarray}
In Eq.~(\ref{V_C}) we have expanded the expression with respect to $\beta(\tau)\sim 1/c $.
Then rather than Eq.~(\ref{DRate}), we will have the following  differential ionization rate
\begin{eqnarray}
\frac{dW}{d^3\textbf{P}d^3\textbf{p}}&=&\frac{4}{\pi} \sum_n \frac{\alpha^2\kappa^3}{(\kappa^2+p^2)^2}\delta \left( \textbf{P}_{\bot} \right) \delta \left( P_{z}+ \frac{n\omega-\sigma_0}{c}\right) \nonumber\\
&\times & \delta \left(\Delta\right) \,J_n^2(\zeta)\left[ 1-\frac{4p_z}{\kappa^2+p^2}\left(\beta_0+\frac{n}{\zeta}  \beta_1\right)\right]\,. \label{DRateC}
\end{eqnarray}
The ionization differential rate  integrated over the ion momenta reads
\begin{eqnarray}
\frac{dW}{d^3\mathbf{p}_e}&\approx & \frac{4}{\pi} \frac{J_n^2(\zeta)}{1+\dfrac{p_{ez}}{m_ic}-\dfrac{m_e}{m_i} \nu}  \label{RateCe}\\
&\times &\frac{\alpha^2\kappa^3}{(\kappa^2+p^2)^2}\left[ 1-\frac{4p_z}{\kappa^2+p^2}\left(\beta_0+\frac{n}{\zeta}  \beta_1\right)\right]\,,\nonumber
\end{eqnarray}
where $p^2=p_{e\bot}^2+\left( p_{ez}-m_e^2c\nu/M\right)^2$ and $p_z=p_{ez}-m_e^2c\nu/M$. The parameters $\zeta,\,\nu,\, n$ are determined by Eqs.~(\ref{zeta}), (\ref{nu}) and (\ref{nnn}).

\begin{figure}
\begin{center}
\includegraphics[width=0.5\textwidth]{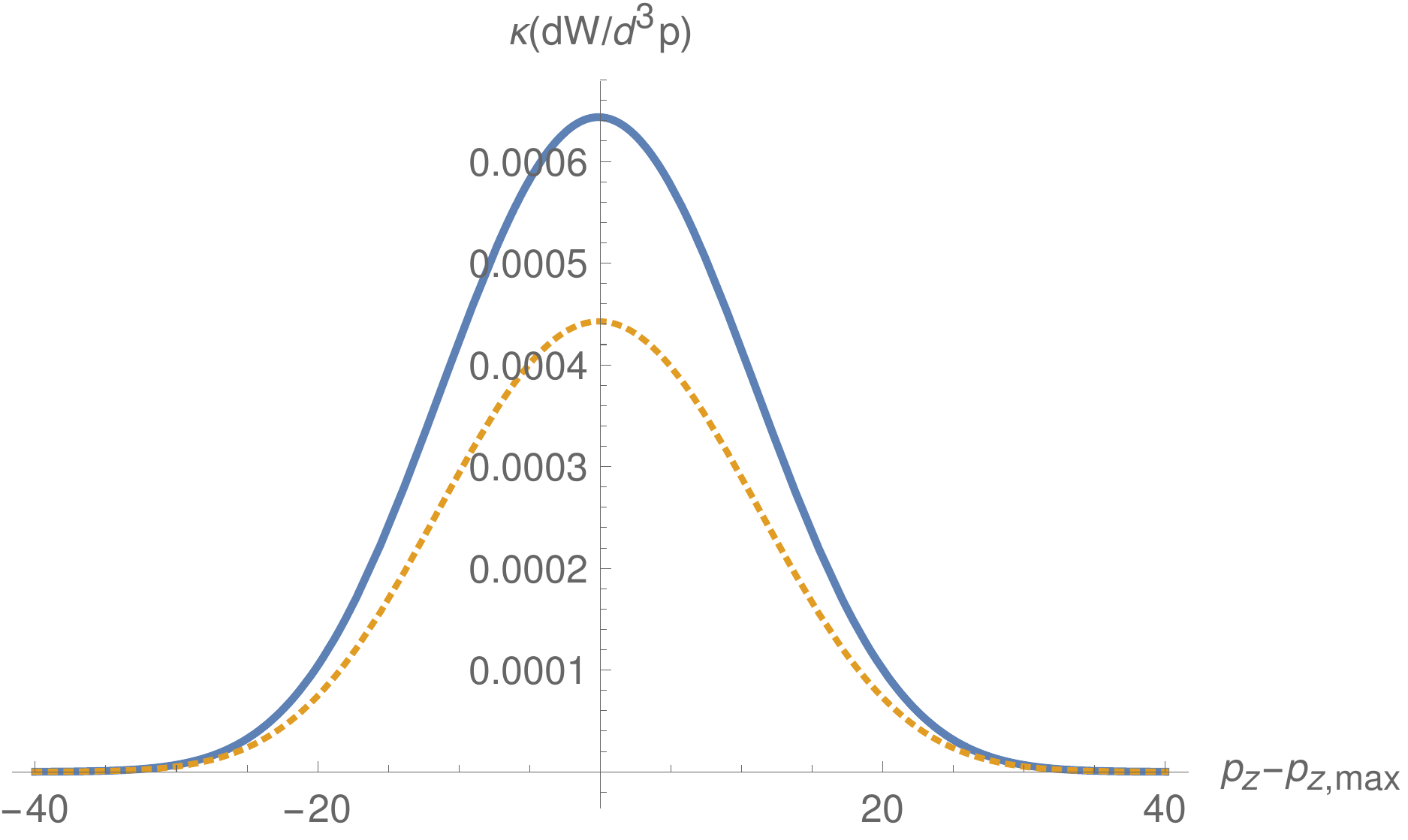}
\end{center}
\caption{\label{fig:1} (color online) The electron distribution function over the momentum along the laser propagation direction $p_z\equiv p_{e\,z}/(m_ec\xi^2/2)$ in the case of a positronium ionization: (solid) via Eq. (\ref{RateCe}) using Coulomb atomic potential; (dashed) via Eq. (\ref{Rate2}) (multipled by a factor of 100) when atomic potential is modelled by a short range potential. The transverse momentum is evaluated at the maximum of the momentum distribution $p_{e\bot}=m_ec\xi (1+\gamma^2/12)$ according to Eq.~(\ref{PsTR}), $p_{z,\, \rm max}\equiv (p_{e\,\bot}/mc\xi)^2+\gamma^2/2$ corresponds to the longitudinal component of the electron momentum at the maximum of the distribution according to Eq.~(\ref{Ps_Z}). The laser parameters are $E_0/E_a=0.2$, $\omega=0.05$ ($\gamma=0.5$, $\xi=0.0172$).}
\end{figure}

The qualitative behaviour of the ionization rate is illustrated in Fig.~\ref{fig:1} on the example of the positronium ionization. The accounting of the exact atomic potential corrects only the value for the ionization rate, but the position of the peak of the momentum distribution is determined by the Bessel function. Therefore, one can rely on the conclusions on the momentum sharing between the ion and the electron presented in this section above. Figure 1 illustrates that  our analytical expression o Eq.~(\ref{Ps_Z}) [which is the particular $m_e=m_i$ case of the general Eq.~(\ref{PeZ})] provides the correct value for the longitudinal component of the electron momentum at the maximum of the distribution.

\section{Conclusion}\label{conclusion}

We have investigated the momentum partition between the constituents of exotic atoms during strong field tunneling ionization. The momentum distribution is deviated from the prediction of the simpleman model. One reason for the deviation is that the electron appears in the continuum with nonvanishing momentum along the laser propagation direction which is due to the effect of the magnetically induced Lorentz force during the under-the-barrier dynamics and due to nonadiabatic effects. The second reason is the impact of the recoil of the atomic core on the tunneling dynamics and, therefore, on the momentum shift of the electron (muon) along the laser propagation direction. The second factor is negligible for common atoms but significant in the case of exotic atoms such as muonic hydrogen and positronium.

\section*{Acknowledgment}

We acknowledge helpful discussions with Prof. C. H. Keitel. DC acknowledges the hospitality of the Max Planck Institute for Nuclear Physics during his three-months visit and thanks Dr. P. P. Corso for funding the Heidelberg visit.

\bibliography{strong_fields_bibliography}

\end{document}